\documentclass{PoS}
\usepackage{psfig}
\usepackage{epsf}

\title{Next-generation astronomy}

\ShortTitle{Next-generation astronomy}

\author{\speaker{Ray P. Norris}\\
        CSIRO Astronomy \& Space Science, PO Box 76, Epping NSW 1710, Australia\\
        E-mail: \email{Ray.Norris@csiro.au}}

\abstract{Fundamental changes are taking place in the way we do astronomy. In twenty years time, it is likely that most astronomers will never go near a cutting-edge telescope, which will be much more efficiently operated in service mode. They will rarely analyse data, since all the leading-edge telescopes will have pipeline processors. And rather than competing to observe a particularly interesting object, astronomers will more commonly group together in large consortia to observe massive chunks of the sky in carefully designed surveys, generating petabytes of data daily. 

We can imagine that astronomical productivity will be higher than at any previous time. PhD students will mine enormous survey databases using sophisticated tools, cross-correlating different wavelength data over vast areas, and producing front-line astronomy results within months of starting their PhD.  The expertise that now goes into planning an observation will instead be devoted to planning a foray into the databases. In effect, people will plan observations to use the Virtual Observatory. Here I examine  the process of astronomical discovery, take a crystal ball to see how it might change over the next twenty years,  and identify further opportunities for the future, as well as identifying pitfalls against which we must remain vigilant.}

\FullConference{Accelerating the Rate of Astronomical Discovery, sps5\\
		August 11-14, 2009\\
		Rio de Janeiro, Brazil}

\begin{document}
\section{Introduction}
Fundamental changes are taking place in the way we do astronomy, with a trend to larger facilities with pipeline processors and wide availability of data,
which should result in a significant increase in astronomical productivity. 

If we extrapolate from current trends, in twenty years time, most astronomers will never go near a cutting-edge telescope, which will be much more efficiently operated in service mode. They will rarely analyse data, since all the leading-edge telescopes will have pipeline processors. And rather than competing to observe a particularly interesting object, astronomers will more commonly group together in large consortia to observe massive chunks of the sky in carefully designed surveys, generating petabytes of data daily. 

We can imagine that astronomical productivity will be higher than at any previous time. PhD students will mine enormous survey databases using sophisticated tools, cross-correlating different wavelength data over vast areas, and producing front-line astronomy results within months of starting their PhD.  The expertise that now goes into planning an observation will instead be devoted to planning a foray into the databases. Increasingly,   "observations" will become virtual, mining data on disks rather than the data in the sky.  



 \vspace{0mm}
\begin{figure}[hbt]
\begin{center}
\epsfysize=80mm
\epsfbox[0 10 1000 600]{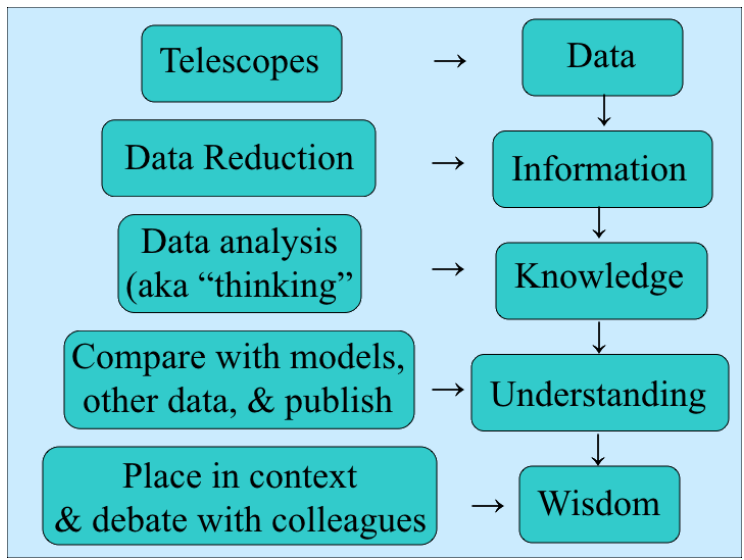}
\caption{The astronomical process, showing how data is transformed into knowledge, understanding, and (occasionally) wisdom. For example, data are transformed into information by the process of data reduction, that information is turned into astronomical knowledge by the process of data analysis, and so on. Missing from this figure are important processes, such as education, outreach, and construction of next-generation telescopes, which are discussed in the text.}
\end{center}
\end{figure}

Occasionally, people confuse data with information, or even knowledge. The difference between them is shown on the right hand side of Fig.1, which illustrates how data can generate information, information can generate knowledge, knowledge can generate understanding, and understanding can generate wisdom. The left hand half of the diagram shows how this is implemented in the field of observational astronomy. 

In this paper, I consider the the effect of technological change on this process of  astronomical discovery. Each section of this paper considers one of these steps in turn, and includes those elements which are missing from this diagram: education, outreach, and the construction of next-generation telescopes. Finally I identify the dangers which accompany these massive changes.

\section{Telescopes and Data}
\subsection{Does size matter?}
Harwit \cite{harwit} has argued that astronomical discoveries tend to be made when new technology enables the construction of a new telescope or instrument that can make observations that were previously impossible . This can be likened to an observational phase space whose axes are the various measurable parameters (e.g. sensitivity, spatial resolution, variability timescale, etc.). Any new observation extending into unexplored phase space is likely to make new discoveries. 

One of the principal axes of the observational phase space is sensitivity, which broadly translates into telescope size. Astronomers planning and building new telescopes tend to be obsessed with size, and some of our most ambitious planned telescopes have names such as "Extremely Large Telescopes" (ELTs) and the "Square Kilometre Array" (SKA) \footnote{Speculation as to whether this is connected to the gender of most astronomers is outside the scope of this paper, but as the gender balance improves, it will be interesting to look for a reduction in the preoccupation with size.}. 

Next-generation telescopes currently being planned or constructed, such as ELTs and the SKA, typically cost billions of dollars, and are already increasingly being built by groups of nations rather than individual nations. To build even bigger telescopes in twenty years time will cost a significant fraction of a nation's wealth, and it is unlikely that any nation, or group of nations,  will set a sufficiently high priority on astronomy to fund such an instrument. So astronomy may be reaching the maximum size of telescope that can reasonably be built.

However, many astronomical discoveries have occurred along other axes of the observational phase space. For example, the discovery of pulsars \cite{burnell} can be attributed to (a) a new instrument sampling an unexplored region of the time-resolution axis, and (b) an astronomer sufficiently familiar with her instrument to recognise an unexpected observational result. So reaching the maximum telescope size should not signal the end of astronomical growth, but merely the end of astronomical inflation, and the need to apply creative thinking to telescope design rather than the brute-force approach of making bigger telescopes. 

For example, thirty years ago optical astronomy was revolutionised by the introduction of CCDs which replaced single-pixel photometers. Radio-astronomers are now arguably witnessing a revolution of comparable scale, with the introduction of phased-array feeds. Their impact is shown in Figure 2, which shows how a previously inaccessible area of observational phase space will  be opened up by the use of phased-array feeds.

In addition, future telescopes have plenty of room to expand along other axes. For example, the time domain is a particularly rich seam that has not yet been adequately tapped, and time-domain astronomy is likely to be increasingly important in the future (\cite{djorgovski}). 

 \vspace{0mm}

\begin{figure}[hbt]
\begin{center}
\epsfysize=90mm
\epsfbox[0 10 1000 400]{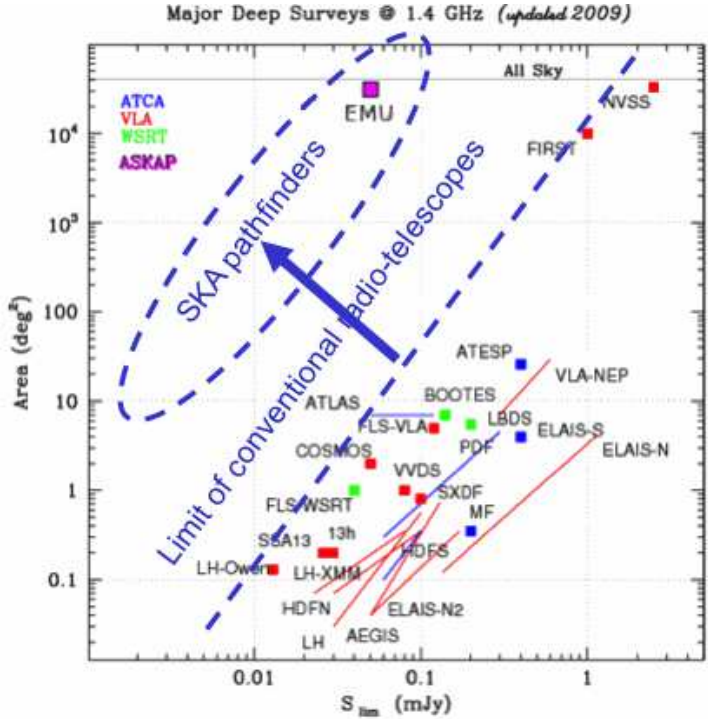}

\caption{The observational phase space of 20cm radio continuum surveys, adapted from \cite{norris}, and originally generated by Isabella Prandoni.  The most sensitive surveys are on the bottom left, and the largest area surveys on the top right. The dashed diagonal line shows the limiting envelope of existing surveys, which are largely limited by available telescope time. Next-generation SKA pathfinders such as the Australian SKA Pathfinder (ASKAP) are able to survey the unexplored region of phase space to the top left by using innovative approaches such as focal plane arrays.}
\end{center}
\end{figure}

\subsection{Surveys and the Virtual Observatory}
Twenty years ago, an astronomer wishing to understand a new type of object would apply for observing time on many different telescopes operating at different wavelengths. To an increasing extent, astronomers can now access data from one of the many large-scale surveys at a depth and resolution comparable to that obtained  from a conventional observation on a large telescope.

In twenty years time, nearly all the sky will have been imaged deeply and catalogued by telescopes and surveys such as WISE \cite{wise} and ASKAP-EMU\cite{norris}, and so an astronomer will be able to access deep multi-wavelength data on any object being studied. Even more importantly, the astronomer will be able to access large numbers of such objects. As a result, astronomical discoveries will not only come from new telescopes observing new regions of observational phase space, but also from astronomers who mine  the databases in innovative ways, and stack images to unprecedented sensitivities. They will effectively be using the databases to form a new axis of observational phase space. The insight and creativity which currently leads to the best observing proposals will, in twenty years time, be focussing on innovative ways of extracting data from the databases. In a sense, we will still be observing, but our observations will increasingly be of a virtual sky on disk rather than the real sky.

For this to happen, we need tools to enable it. The Virtual Observatory aims to provide those tools, and although those tools have been slower to arrive than some of us would like, they are starting to make an impact in front-line research, and this trend will increase as the large surveys become more widely accessible. But we're not there yet, and only the simplest types of query are currently supported.  In the future, creative users will design  tools which handle  much more complex and innovative queries to address specific astrophysical questions. The ingenuity that at present is applied to designing new telescopes and instruments will in future need to be applied to the challenge of designing new "instruments" (software tools) for the Virtual Observatory.

\section{Data Reduction and Information}
At present, most reduction of astronomical data (by which I include calibration and editing) is carried out by humans, and most is immensely repetitive, representing an enormous waste of high-calibre human resources. In defence of this position, occasionally astronomical data contains unexpected glitches, which might either be instrumental errors, requiring the data to be flagged or discarded, or it might be an unexpected discovery. Pulsars may not have been discovered if a computer had been reducing Jocelyn Bell Burnell's data. On the other hand, such discoveries are rare, and most students will make more discoveries by spending their time learning about astrophysics than by reducing data.

In any case, a telescope producing petabytes of data will make it impossible for users to process data. Almost all next-generation telescopes now being designed offer automated pipeline reduction of data. The user will log in, obtain their calibrated images, cubes, or catalogs, and the time previously spent reducing and calibrating data will increasingly be used to analyse and interpret the data. 

Other benefits of automated data reduction include
\begin{itemize}
\item it is objective and reproducible,
\item it should incorporate "best practice" data reduction, which may represent a significant advance over techniques learnt some years ago by the average user, 
\item it enables new students, or theorists, or researchers experienced at a different wavelength, to access the capabilities without having to learn arcane knowledge, thus encouraging cross-fertilisation,
\item it enables people in developing countries to access cutting-edge  capabilities. 
\end{itemize}

Another trend is that data are increasingly being placed in the public domain. Whilst proprietary periods are necessary in some cases to motivate and protect the researchers who spend years designing and building an instrument,  it is clear that much astronomical discovery occurs after the data are released to other groups, who are able to add value to the data by combining it with data. models, or ideas which may not have been accessible to the instrument designers. 

As large-scale surveys in the public domain become more common, a future observing run may take the form of
\begin{itemize}
\item Select what sources (or parameters) you want to observe
\item Observe them by mining survey databases
\item Perhaps think of a really clever way of exploring a new part of observational phase-space using the database
\item Make new discoveries by combining data from different wavelengths/regions of parameter space 
\end{itemize}

Of course, there is still a role for major observational facilities: not only to produce survey data, but also to probe deeply into small areas of space, going well beyond the available survey data on particularly interesting objects or parts of parameter space.

Incentives need to be provided to place data in the public domain. At present, astronomers' career prospects often depend critically on the number of publications, or citations, and do not depend at all on data they have obtained and placed in the public domain to help others make discoveries. This can be remedied by attaching a tag, such as a Digital Object Identifier, to a dataset, or to  a particular view or compilation of data. This would work as follows:

\begin{enumerate}

\item Every time an author publishes a paper in a refereed journal, her catalogs, images, or results appear in the data centres within days.

\item Papers include Digital Object Identifiers for the data they use, making it easy to track who's using what data. So an astronomer's CV lists citations for her data as well as those for her publications.

\item As a result, an astronomer places her data in the public domain at the earliest opportunity, thereby increasing her citations, and accelerating the rate of astronomical discovery even further.

\item Every image or result derived from public-domain databases is accompanied in the published paper by a digital identifier. The reader can click on the identifier, and he can see the raw (or processed) data from which the author obtained her results. He can verify her results, tweak the analysis, perhaps even improve on her result.
\end{enumerate}

\section{Data Analysis and Knowledge}

Just as the volumes of raw data become too large to manage comfortably by an individual, so too do the volumes of processed information. For example, in 2013, EMU \cite{norris} will discover and catalogue about 70 million galaxies at radio wavelengths. The science will only flow from these catalogues when we cross-identify them with optical/infrared objects from surveys such as WISE \cite{wise}, VISTA \cite{vista}, Sloan \cite{sloan}, and SkyMapper \cite{skymapper}. Perhaps 70\% of the cross-identifications can be performed automatically with a simple algorithm which compares catalogs for simple isolated objects. For extended or multiple objects, which is where much of the best science is likely to lie, the process will be much harder, and at present we do not have algorithms that can handle these hard questions. And even the largest team of PhD students will not be able to handle tens of millions of cross-identifications by eye.

Projects such as GalaxyZoo \cite{zoo} already tackle a similar problem, that of galaxy classification from the Sloan survey, by employing enthusiastic citizens from around the world who would like to contribute to the scientific endeavour. Users are led through a simple decision tree to make a complex classification of a galaxy. Even an untrained human brain is far superior to the best algorithm currently available, and Galaxy Zoo success rates are high, with real science being generated.


We expect such initiatives to increase in the future, perhaps even increasing the intellectual demands made on a subset of users, so that these projects may tap into the brightest minds on the planet, many of whom do not have the educational background or opportunities to participate in conventional science. We know that potential Einsteins exist in third-world villages, and projects such as GalaxyZoo and its descendants will enable these people to contribute to human knowledge. More selfishly, such projects enable astronomy to tap into vast reserves of raw untapped brainpower.

\section{Publishing and Understanding}
The knowledge gained from an observation does not deliver any significant  value to the wider community until it is published. 

It is easy to point out flaws in our current peer-reviewed journal system:
\begin{itemize}
\item While many referees add real value to the process, others are sloppy or obstructive. 
\item It is far easier to publish a mediocre but correct paper than a truly innovative paper which challenges deeply-held beliefs. 
\item Papers published in expensive journals are not available to students in impoverished Universities. 
\end{itemize}
On the other hand, the peer-review system protects us from a morass of junk papers. It may not be a perfect system, but it's probably the best we have at present. 

Some deficiencies of the system are overcome by systems such as ArXiv \cite{astroph} which offer fast informal publication of preprints, while other open-access journals offer alternative modes of publication. Other systems of peer review are under trial in other disciplines, include unrefereed publication on the web accompanied by comments from peers and readers. The science community have only just started to explore the capabilities of Web 2.0 - it is likely that social networking sites have much to offer the process of dissemination of science, but we have not yet tapped into it in any significant way.

A discussion on astronomical publishing would be incomplete without mention of ADS \cite{ads} which has become the main entry point into the literature for most astronomers. While we may speculate on what advances may accelerate the rate of astronomical discovery in the next twenty years, it is clear that ADS has significantly advanced the the rate of astronomical discovery compared to twenty years ago.

We don't yet have the perfect system, and here I am not advocating one system against another. The important thing is to continue experimenting with new models until we find a better process. In twenty years time, the publication process may be quite different. 

\section{Education and Outreach}

Many astronomers view outreach as something that we "ought" to do, to compensate the government, or the broader community, for the billions of dollars poured into astronomy. Here I take a much more selfish view. If we truly want to accelerate the rate of astronomical discovery, we need to invest a significant fraction of our effort into outreach to (a) demonstrate to the community that we are doing something worthwhile, so that they continue to support us, (b) search for opportunities for astronomy to act as a positive social force (e.g. build cross-cultural bridges, attract young people into science and technology), and (c) attract the brightest young minds into astronomy.

Conventional outreach (media releases, TV programs, public talks, high-school programs) will obviously continue to be important. But we must also look for innovative ways to use new technologies. SETI@home and GalaxyZoo have already shown two ways to engage thousands of enquiring minds across the planet. Other individuals conduct astronomy classes in Second Life. It is likely that we have hardly started to tap the possibilities, and, as in publishing, the important thing is to keep experimenting.

\section{Wisdom}

In twenty years time, I expect to be able to click on an object in a paper, and see its image at all wavelengths. When I publish a paper, I expect the results  to be immediately available in the data centres used by the Virtual Observatory. I expect to be able to survey vast swathes of the Universe in a miriad of different wavelengths and parameters, and if I fail to understand the complex processes driving its evolution, then I have only my brain to blame, not the lack of data or information.

I expect to see astronomical discoveries being made at an astonishingly increasing rate, and I expect to have a tool which intelligently filters out the papers that I want to read, to keep abreast of developments in my field. Soon after starting their PhD, my PhD students will be able to mine vast astronomical databases using sophisticated tools, trying out "what-if"s as part of their education, and discovering cosmic trends for themselves. The best of them may even produce cutting-edge results within months of starting their PhD. It will truly be a Golden Age of Astronomy!

On the other hand, few if any of my students will ever have observed on a major telescope, and only the brightest will really understand how a telescope works. Most will fail to recognise artefacts in the data, and so the literature will be peppered with spurious results. Sometimes a truly astounding discovery will be dismissed as a "bit of scruff" in the data. So maybe our Golden Age will contain some Pyrites.

And when my students graduate, and eventually become fully-fledged astronomers with budgets and responsibilities, their lack of familiarity with instrumentation may prevent them from thinking creatively about how to design next-generation telescopes. So we must make the most of our Golden Age, because it may never be upgraded to Platinum.

We must embrace technology, and never be afraid to experiment with new ways of doing things. At the same time, we must pay attention to the process of doing astronomy, and ensure that our technological systems support, rather than obstruct, the fragile creative processes which ultimately lead to new discoveries.

\clearpage

\end{document}